\documentstyle[aps,preprint]{revtex}

\begin{document}
\draft
\tightenlines
\title
{\bf Ginzburg-Landau Expansion in a Toy Model of Superconductor
with Pseudogap}
\author{A.I.Posazhennikova,\ M.V.Sadovskii}
\address
{Institute for Electrophysics,\\ Russian Academy of Sciences,\
Ural Branch, \\ Ekaterinburg,\ 620049,\ Russia\\
E-mail:\ posazhen@ief.uran.ru,\ sadovski@ief.uran.ru} 
%\date{} 
\maketitle

\begin{center}
{\sl JETP{\bf 115}, No.2 (1999)}\\
{\sl cond-mat/9806199}
\end{center}

\begin{abstract}
We propose a toy model of electronic spectrum of two-dimensional system with
``hot-patches'' on the Fermi surface, which leads to essential renormalization
of spectral density (pseudogap). Within this model we derive Ginzburg-Landau
expansion for both $s$-wave and $d$-wave Cooper pairing and analyze the
influence of pseudogap formation on the basic properties of superconductors.
\end{abstract}

\newpage
\narrowtext
\section{Introduction}

Among the number of anomalies of electronic properties of high-temperature
superconductors particularly interesting is the observation of the pseudogap
in the spectrum of elementary excitations in the underdoped region of the
phase diagram\cite{Ran,RC}. Especially striking evidence of this unusual state
were obtained in angle-resolved photoemission experiments on 
$BSCO$\cite{Ding,D}, which demonstrated the presence of significantly
anisotropic changes in carrier spectral density in normal phase ($T>T_c$). 
In these experiments the maximal value of the pseudogap was observed close to
the point ($\pi,0$) of the Brillouin zone, while in the diagonal direction
pseudogap changes were absent. Accordingly, in the vicinity of the point
($\pi,0$) the Fermi surface is practically completely destroyed, while it is
conserved in the region around the zone diagonal. This picture is usually
described as ``$d$-wave'' symmetry of the pseudogap, which is the same as the
symmetry of superconducting gap in these systems. These anomalies persist up
to the temperatures $T\simeq T^*$ which are much larger than $T_c$.

There is a number of theoretical approaches aiming to explain these
anomalies. Among these, two main directions of thought can be described as
the picture of Cooper pairs formation at temperatures much higher than
$T_c$ \cite{Ran,Gesh,EK}, and alternative scheme assuming the decisive role
of fluctuations of antiferromagnetic short-range order 
\cite{Kam,Bar,Pines,Sch,SchP}. 

Most of theoretical efforts up to now were dedicated to the studies of pseudogap
state of HTSC in normal phase,\ i.e. for $T>T_c$. The aim of the present work
is qualitative study of the influence of the pseudogap in electronic spectrum
on the basic superconducting properties. We assume the validity of the ideology
of antiferromagnetic fluctuations\cite{Kam,Bar,Pines,Sch,SchP}, but introduce
a very simplified model of the pseudogap state in normal phase, which allows
the complete analysis in analytic form. Within this model we shall perform
microscopic derivation of Ginzburg--Landau expansion for systems with both
$s$-wave and $d$-wave pairing and analyze the qualitative effects of
pseudogap formation (destruction of the parts of the Fermi surface) on basic
superconducting properties.

\section{Elementary Model for a Pseudogap State of Two-Dimensional Electronic
System.}

As mentioned above, we propose a greatly simplified model of the pseudogap state
based upon the picture of well developed fluctuations of antiferromagnetic
short-range order, which is close to the model of ``hot-spots'' on the Fermi
surface \cite{Sch,SchP}. Let us assume that the Fermi surface of two-dimensional
electronic system has the form shown in Fig.1. Analogous form of the Fermi
surface was proposed in Ref. \cite{Z}, where it was stressed that this form is
very similar to that observed in a number of HTSC-systems \cite{Dessau,Shen}.
We shall assume also that fluctuations of short-range order are static and
Gaussian and define the appropriate correlation function as (Cf.Ref.\cite{Kam}):  
\begin{equation}
S({\bf q})=\frac{1}{\pi^2}\frac{\xi^{-1}}{(q_x-Q_x)^2+\xi^{-2}}
\frac{\xi^{-1}}{(q_y-Q_y)^2+\xi^{-2}}
\label{fluct}
\end{equation}
where 
$\xi$-is the correlation length of these fluctuations. 
Here we take either
$Q_x=\pm 2k_F$ and $Q_y=0$ or $Q_y=\pm 2k_F$ and $Q_x=0$. 
We shall assume that only electrons from flat ``hot'' patches on the Fermi
surface shown in Fig.1 interact with these fluctuations.
Effective interaction of electrons with these fluctuations
will be described as $(2\pi)^2W^2S({\bf q})$, where the parameter $W$ has the
dimension of energy and defines the characteristic energy scale (width) of the
pseudogap.
\footnote {More formally we introduce here an effective
interaction ``constant'' of electrons with fluctuations:

$W_{\bf p}=W[\theta(p_x^0-p_x)\theta(p_x^0+p_x)+
\theta(p_y^0-p_y)\theta(p_y^0+p_y)]$.} 
It is clear that in our model this scattering is of purely 
one-dimensional nature.
The choice of scattering vector ${\bf Q}=(\pm 2k_F,0)$ or ${\bf Q}=(0,\pm 2k_F)$
corresponds in general 
to the picture of incommensurate fluctuations. Below we shall also consider a
special case of commensurate scattering with
${\bf Q}=(\frac{\pi}{a},\frac{\pi}{a})$ (where $a$ is lattice spacing).
In the limit of $\xi\rightarrow \infty$ our model can be solved exactly 
using methods proposed rather long ago in Refs.\cite{C1,C2}, 
while in case of finite $\xi$ (although with some caution \cite{Sch,SchP,W}) 
by the method of Refs.\cite{C79,C91}. Below we shall only consider the toy
case of $\xi\rightarrow \infty$, when the effective interaction
with fluctuations (\ref{fluct}) takes the following oversimplified form:  
\begin{equation} 
(2\pi)^2W^2\left\{\delta(q_x\pm 2k_F)\delta(q_y)+\delta(q_y\pm 
2k_F)\delta(q_x)\right\} 
\label{WW} 
\end{equation} 
In this case it is possible to make 
the complete summation of all perturbation series for the electron scattered 
by these fluctuations \cite{C1,C2} and obtain the one-electron Green's 
function in the following form:  
\begin{equation} G(\epsilon_n,p)=\int 
\limits_0^\infty d{\zeta}e^{-\zeta} 
\frac{i\epsilon_n+\xi_p}{(i\epsilon_n)^2-\xi_p^2-\zeta W(\phi)^2},
\label{fgrina}
\end{equation}
where $\xi_p=v_F(|{\bf p}|-p_F)$ ($v_F$- Fermi velocity), 
$\epsilon_n=(2n+1)\pi T$, while $W(\phi)$ is defined for
$0\leq\phi\leq\frac{\pi}{2}$ as:
\begin{equation}
W(\phi)=\left\{
\begin{array}{ll}
W & ,0\leq\phi\leq\alpha,\>\frac{\pi}{2}-\alpha\leq\phi\leq\frac{\pi}{2} \\
0 & ,\alpha\leq\phi\leq\frac{\pi}{2}-\alpha
\end{array}
\right.
\label{w}
\end{equation}
where $\alpha=arctg(\frac{p_y^0}{k_F})$, $\phi$ - polar angle, defining
the direction of ${\bf p}$ in ($p_x,p_y$) plane. For other values of $\phi$  
we can define $W(\phi)$ analogously to (\ref{w}) from the obvious symmetry.
It is easily seen that changing $\alpha$ within the interval 
$0\leq\alpha\leq\frac{\pi}{4}$, we in fact change the size of 
``hot patches'' on the Fermi surface, where the ``nesting'' condition
$\xi_{p-Q}=-\xi_p$ is satisfied. In particular, $\alpha=\pi/4$ corresponds to
the square Fermi surface, where the ``nesting'' condition is satisfied
everywhere. Outside the ``hot patches'' (second inequality in (\ref{w})) the
Green's function (\ref{fgrina}) is simply a free electron one.

Spectral density corresponding to Green's function (\ref{fgrina}) is given by:
\begin{eqnarray}
\rho(\epsilon\xi_p)=
-\frac{1}{\pi}sign\epsilon Im G(\epsilon \xi_p)=\\
\left\{
\begin{array}{lcl}
\frac{1}{W^2}(|\epsilon|+\xi_psign\epsilon)\theta(\epsilon^2-\xi_p^2)
exp^{\frac{\epsilon^2-\xi_p^2}{W^2}} & , \mbox{ for } &  
0\leq \phi\leq\alpha, \frac{\pi}{2}-\alpha\leq\phi\leq\frac{\pi}{2} \\
\delta(\epsilon-\xi_p) & , \mbox{ for } & \alpha\leq\phi\leq\frac{\pi}{2}-\alpha
\end{array}
\right.
\label{ro}
\end{eqnarray}
and similarly for the other quadrants of the Brillouin zone.
This expression (\ref{ro}) demonstrates non-Fermi-liquid (pseudogap) behavior
with ``$d$-wave'' symmetry on the ``hot patches'' and free-like Fermi-liquid
behavior on ``cold patches'' of the Fermi surface. Graphically the behavior of
the spectral density on the ``hot patch'' is shown in Fig.2.
Taking into account that the integral over the polar angle $\phi$ of the 
arbitrary function of $f(W(\phi))$ (where $W(\phi)$ is defined in (\ref{w})) 
is obviously given by:
\begin{equation}
\int_{0}^{2\pi}d{\phi}f(W(\phi))=8\alpha f(W(\phi))+(2\pi-8\alpha)f(0),
\label{int}
\end{equation}
we can easily find from (\ref{ro}) the density of states:
\begin{equation}
\frac{N(E)}{N_0(0)}=-\frac{1}{\pi}\int \limits_0^{2\pi}\frac{d\phi}{2\pi}
\int \limits_{\infty}^{\infty}d{\xi_p} Im G^R(\epsilon\xi_p)=
4\alpha/\pi N_W(\epsilon)+(1-4\alpha/\pi)N_0(0)
\label{NN}
\end{equation}
where $N_0(0)$ - is the density of states of free electrons at the Fermi level, 
while $N_W(\epsilon)$ - is the density of states of the one-dimensional problem
(square Fermi surface), which was derived earlier in Refs.\cite{C1,C2}:
\begin{equation}
\frac{N_W(\epsilon)}{N_0(\epsilon)}=\left|\frac{\epsilon}{W}\right|
\int\limits_0^\frac{\epsilon^2}{W^2}d{\zeta}\frac{e^{-\zeta}}
{\sqrt{\frac{\epsilon^2}{W^2}-\zeta}}=
2\left|\frac{\epsilon}{W}\right|
exp(-\frac{\epsilon^2}{W^2})Erfi(\frac{\epsilon}{W})
\label{NW}
\end{equation}
where $Erfi(x)$- is probability integral of imaginary argument.

In Fig.3 we show the graphic dependencies of the density of states in our model
for different values of parameter $\alpha$, i.e. for ``hot patches'' of
different sizes. It is seen that the pseudogap in the density of states is
rather rapidly smeared with diminishing size of the ``hot patches'' and in
general is not very pronounced. In some sense the effect of diminishing
$\alpha$ is similar to that of diminishing correlation length of fluctuations 
$\xi$ \cite{C79,C91}, so that our approximation of $\xi\rightarrow\infty$, 
is probably not a great limitation of our model. The advantage of this
approximation on the other hand is in possibility to obtain all the major
results in analytic form.

In conclusion of this section we present a short discussion of the case of
commensurate fluctuations with ${\bf Q}=(\frac{\pi}{a},\frac{\pi}{a})$. 
In Fig.4 we show the model Fermi surface used in this case. The pseudogap is
opened in the direction of diagonals in the Brillouin zone, which contradicts
experiments on HTSC-systems, but this case is interesting from purely
theoretical point. The model can be solved analogously to the previous case and
generalizes the one-dimensional solution first found in Ref.\cite{W}.  
One-electron Green's function has the form similar to (\ref{fgrina}), with
$W(\phi)$ again a function of $\phi$ with period $\pi/2$ , but ``rotated'' 
with respect to the previous case by an angle of $\pi/4$. For the interval of 
$-\pi/4+\alpha\leq\phi\leq\pi/4+\alpha$ we have:
\begin{equation}
W(\phi)=\left\{
\begin{array}{ll}
W & ,\pi/4-\alpha\leq\phi\leq\pi/4+\alpha \\
0 & ,-\pi/4+\alpha\leq\phi\leq\pi/4-\alpha
\end{array}
\right.
\label{Wphi}
\end{equation}
where $0\leq\alpha\leq\pi/4$. Also in this case we must take into account
a different combinatorics of Feynman diagrams, corresponding to the
scattering of electrons on commensurate fluctuations \cite{W}. As a result
in Eq.(\ref{fgrina}) we have to replace the integral
\begin{equation}
\int_{0}^{\infty}d{\zeta}e^{-\zeta}
\label{intg}
\end{equation}
by
\begin{equation}
\int_{0}^{\infty}d{\zeta}\frac{1}{2\sqrt{\pi\zeta}}e^{-\zeta/4}.
\label{integ}
\end{equation}

\section{Equation for $T_{c}$.}

Let us now consider the case of superconducting pairing in our model.
Assume the usual separable form of pairing interaction \cite{PS}:
\begin{equation}
V({\bf p,p'})=V(\phi,\phi')=-Ve(\phi)e(\phi'),
\label{VV}
\end{equation}
where $\phi$ - is again an angle, determining the direction of electronic
momentum ${\bf p}$ in the plane, and for $e(\phi)$ we assume the simple
model dependence: 
\begin{equation}
e(\phi)=
\left\{
\begin{array}{ll}
1 &,(\mbox{$s$-wave pairing})\\ 
\sqrt{2}cos(2\phi) &,(\mbox{$d$-wave pairing})
\end{array}.
\right.
\label{ephi}
\end{equation}
Interaction constant of attractive interaction $V$ is assumed as usual to be
non-zero in some region of the width of $2\omega_c$ around the Fermi level
($\omega_c$ - is characteristic frequency of quanta, responsible for
attractive interaction). In this case the superconducting gap (order parameter) 
takes the following form:
\begin{equation}
\Delta({\bf p})\equiv \Delta(\phi)=\Delta e(\phi).
\label{DD}
\end{equation}
 
Equation for superconducting transition temperature $T_c$ is obtained from
the usual equation determining Cooper's instability: 
\begin{equation}
1-\chi(0,0)=0,
\label{chi}
\end{equation}
where the generalized Cooper's susceptibility $\chi(0,0)$ can be calculated by
exact summation of all diagrams, taking into account the scattering on 
fluctuations of short-range order (\ref{WW}), in a way similar to that used in
Refs.\cite{C1,C2} to calculate the polarization operator. 
As a result we obtain the following equation for $T_c$:  
\begin{eqnarray} 
\frac{1}{V}=-\int_{0}^{\infty}d{\zeta}e^{-\zeta}T_{c}\sum_n
\int_{0}^{\infty}\frac{d^2{p}}{(2\pi)^2} e^2(\phi)
\Biggl\{(G_{\zeta W^2}(\epsilon_n;{\bf p,p})
G_{\zeta W^2}(-\epsilon_n;-{\bf p,-p})+ \\
F_{\zeta W^2}(\epsilon_n;{\bf p,p-Q})
F_{\zeta W^2}(-\epsilon_n;-{\bf p,-p+Q}\Biggr\},
\label{urnatc}
\nonumber
\end{eqnarray}
where
\begin{eqnarray}
G_{\zeta W^2}(\epsilon_n;{\bf p,p})=\frac{i\epsilon_n+\xi_p}
{(i\epsilon_n)^2-\xi_p^2-\zeta W^2(\phi)}; \\
F_{\zeta W^2}(\epsilon_n;{\bf p,p-Q})=\frac{\sqrt{\zeta}W(\phi)}
{(i\epsilon_n)^2-\xi_p^2-\zeta W^2(\phi)};
\label{funkgr}
\nonumber
\end{eqnarray}
are appropriately defined ``normal'' and ``anomalous'' Green's functions for
the system with dielectric gap \cite{C1,C2}.

After some standard transformations from (\ref{urnatc}) we get: 
\begin{equation}
\frac{1}{V}=\int_{0}^{\infty}d{\zeta}e^{-\zeta}T_{c}\sum_n\int_{0}^{\infty}
\frac{d^2{p}}{(2\pi)^2}
\frac{e^2(\phi)}{\epsilon_{n}^2+\xi_{p}^2+\zeta W(\phi)^2},
\label{5}
\end{equation}
so that after summation over frequencies we obtain:
\begin{equation}
\frac{1}{V}=\frac{N(0)}{2\pi}\int_{0}^{\infty}d{\zeta}e^{-\zeta}\int_{-\infty}^{\infty}
d{\xi}\int_{0}^{2\pi}d{\phi}\frac{e^2(\phi)}{2\sqrt{\xi^2+\zeta W(\phi)^2}}
th\frac{\sqrt{\xi^2+\zeta W(\phi)^2}}{2T_c}.
\label{6}
\end{equation}

Performing the angular integration over $\phi$ similarly to (\ref{int}), 
for the case of $s$-wave pairing we get: 
\begin{equation}
\frac{1}{g}=\frac{4\alpha}{\pi}\int_{0}^{\infty}d{\zeta}e^{-\zeta}
\int_{0}^{\omega_c}
d{\xi}\frac{1}{\sqrt{\xi^2+\zeta W^2}}
th\frac{\sqrt{\xi^2+\zeta W^2})}{2T_c}+(1-\frac{4\alpha}{\pi})
\int_{0}^{\omega_c}d{\xi}\frac{1}{\xi}th\frac{\xi}{2T_c}.
\label{8}
\end{equation}
while for the case of $d$-wave pairing:
\begin{eqnarray}
\frac{1}{g}=\frac{sin(4\alpha)+4\alpha}{2\pi}\int_{0}^{\infty}d{\zeta}e^{-\zeta}
\int_{0}^{\omega_c}
d{\xi}\frac{1}{\sqrt{\xi^2+\zeta W^2}}
th\frac{\sqrt{\xi^2+\zeta W^2})}{2T_c}+  \nonumber \\
\frac{\pi-4\alpha-sin(4\alpha)}{2\pi}
\int_{0}^{\omega_c}d{\xi}\frac{1}{\xi}th\frac{\xi}{2T_c}.
\label{ddd}
\end{eqnarray}
where $g=N(0)V$ - is dimensionless pairing coupling constant. In Fig.5(a,b) we
show the graphic dependencies of $T_c/T_{c0}$ on the parameter $W/T_{c0}$, 
determining the effective width of the pseudogap, for different values of
$\alpha$ ($T_{c0}$--is superconducting transition temperature for an ``ideal''
system without pseudogap). It is clearly seen that for both types of pairing
the appearance of the pseudogap on ``hot patches'' of the Fermi surface leads
to significant suppression of $T_c$, which becomes stronger with the growing
size of ``hot patches''. Naturally enough for the case of $d$ - wave pairing
this effect of $T_c$ suppression is much stronger than in  $s$ - wave case,
because the dielectrization of electronic spectrum (pseudogap) operates ``out
of phase'' with pairing interaction.

In case of commensurate fluctuations (Fig.4) and $d$-wave pairing the
$T_c$ - equation takes the following form:
\begin{eqnarray}
\frac{1}{g}=\frac{4\alpha-sin(4\alpha)}{2\pi}
\int_{0}^{\infty}d{\zeta}\frac{1}{2\sqrt{\pi\zeta}}e^{-\zeta/4}
\int_{0}^{\omega_c}
d{\xi}\frac{1}{\sqrt{\xi^2+\zeta W^2}}
th\frac{\sqrt{\xi^2+\zeta W^2})}{2T_c}+ \nonumber \\
\frac{\pi-4\alpha+sin(4\alpha)}{2\pi}
\int_{0}^{\omega_c}d{\xi}\frac{1}{\xi}th\frac{\xi}{2T_c}.
\label{8a}
\end{eqnarray}

The appropriate dependencies of $\frac{T_c}{T_{c0}}$ on $\frac{W}{T_c0}$ 
for different values of $\alpha$ in this case are shown in Fig.6.
Here the effect of $T_c$ suppression by the pseudogap is less significant
because the maximum value of superconducting gap is achieved on ``cold
patches'' of the Fermi surface, where pseudogap is absent.

\section{Ginzburg--Landau Expansion.}

Ginzburg--Landau expansion for the difference of free energy density of
superconducting and normal state can be written in usual form:
\begin{equation}
F_{s}-F_{n}=A|\Delta_{q}|^2+q^2 C|\Delta_{q}|^2+\frac{B}{2}|\Delta_{q}|^4,
\label{GL}
\end{equation}
where $\Delta_q$ - is the amplitude of the Fourier component of the order
parameter:
\begin{equation}
\Delta(\phi,q)=\Delta_qe(\phi).
\label{FF}
\end{equation}

Actually (\ref{GL}) is determined by the loop expansion for free energy of
electron scattered by fluctuations of the order parameter with small
wave-vector ${\bf q}$, which is shown in Fig.7, where in all loops we have to
perform exact summation of all scattering processes due to fluctuations of
short range order (\ref{WW}). Again this is easily done by methods of Refs.
\cite{C1,C2}. In other respects all calculations are similar to that done by us
in Ref.\cite{PS}. 
\footnote{Similar in spirit Ginzburg-Landau analysis for Peierls transition
in quais-one-dimensinal systems was performed in Ref.\cite{McK}.}
As in this previous work the subtraction of the second
diagram in Fig.7 guarantees the zero value of the coefficient $A$ at the
transition point $T=T_c$. Finally, the coefficients of Ginzburg--Landau
expansion are written as:
\begin{equation}
A=A_{0}K_{A};\qquad   C=C_{0}K_{C};\qquad
B=B_0K_B,
\end{equation}
where $A_{0}$, $C_{0}$ and $B_0$ define these coefficients for the case of
two-dimensional isotropic $s$-wave superconductor in the absence of pseudogap
($\alpha=0$) :                                             
\begin{equation}
A_{0}=N(0)\frac{T-T_{c}}{T_{c}};\qquad
C_{0}=N(0)\frac{7\zeta(3)}{32\pi^{2}}\frac{v_F^2}{T_c^2};\qquad
B_0=N(0)\frac{7\zeta(3)}{8\pi^{2}T_c^2},
\label{11}
\end{equation}
while all peculiar dependencies characteristic to our model are contained in
dimensionless coefficients $K_{A}$, $K_{C}$ and $K_B$. In the absence of the
pseudogap all these dimensionless coefficients are equal to 1, only in the
case of $D$-wave pairing $K_B=3/2$.

Direct calculations give: 
\begin{equation}
A=N(0)\frac{T-T_c}{2T_c^2}\frac{1}{2\pi}\int_{0}^{\infty}d{\zeta}e^{-\zeta}
\int_{0}^{\omega_c}
d{\xi}\int_{0}^{2\pi}d{\phi}
\frac{e^2(\phi)}{ch^2\frac{\sqrt{\xi^2+\zeta W(\phi)^2}}{2T_c}},
\label{11a}
\end{equation}
so that after the integration over $\phi$ we obtain:
\begin{equation}
K_A=\frac{1}{2T_c}\beta_a
\int_{0}^{\infty}d{\zeta}e^{-\zeta}
\int_{0}^{\omega_c}d{\xi}
\frac{1}{ch^2\frac{\sqrt{\xi^2+\zeta W^2}}{2T_c}}+
1-\beta_a,
\label{KK}
\end{equation}
where 
\begin{equation}
\beta_a=\left\{
\begin{array}{ll}
\frac{4\alpha}{\pi} & ,\mbox{$s$-wave pairing} \\
\frac{4\alpha+sin(4\alpha)}{\pi} & ,\mbox{$d$-wave pairing}.
\end{array}
\right.
\label{betta}
\end{equation}
In Fig.8 we show the dependencies of $K_A$ on the pseudogap width $W/T_{c0}$ 
for different values of $\alpha$. We show these only for $s$-wave case,
for $d$-wave pairing these dependencies are qualitatively similar, but all the
changes of the coefficient take place on much smaller scales of $W/T_{c0}$, 
as in Fig.5.

To calculate $C$ it is necessary to make Taylor expansion in powers of 
${\bf q}$ in
\begin{eqnarray}
-\int_{0}^{\infty}d{\zeta}e^{-\zeta}T_{c}\sum_n
\int_{0}^{\infty}\frac{d^2{p}}{(2\pi)^2}e^2(\phi)
\Biggl\{(G_{\zeta W(\phi)^2}(\epsilon_n;{\bf p_+,p_+})
G_{\zeta W(\phi)^2}(-\epsilon_n;{\bf-p_-,-p_-})+ \\
F_{\zeta W(\phi)^2}(\epsilon_n;{\bf p_+,p_+-Q})
F_{\zeta W(\phi)^2}(-\epsilon_n;{\bf
-p_-,-p_-+Q}\Biggr\},
\label{14}
\nonumber
\end{eqnarray}
where ${\bf p_{\pm}}={\bf p} \pm \frac{{\bf q}}{2}$,
and find the coefficient of ${\bf q^2}$.  To simplify expressions let us
introduce the following notations:
$G_{\zeta W(\phi)^2}(\epsilon_n;{\bf p,p})\equiv G_{pp};\
F_{\zeta W(\phi)^2}(\epsilon_n;{\bf p,p-Q})\equiv F_{pp-Q}$.

After some complicated but purely technical calculations we obtain
$C$ in the following form: 
\begin{equation}
C=-T_c\frac{N(0)}{2\pi}v_F^2\sum_n\int_{0}^{\infty}d{\zeta}e^{-\zeta}
\int d{\xi}\int_{0}^{2\pi} d{\phi}\frac{cos(\phi)^2e^2(\phi)
(\xi^2-3\epsilon_n^2
-3\zeta W(\phi)^2)}{2(\epsilon_n^2+\xi^2+\zeta W(\phi)^2)^3}.
\label{20}
\end{equation}

Accordingly, after the integration over $\xi$ and $\phi$ we get for the
dimensionless coefficient $K_C$:
\begin{equation}
K_C=\beta_c\frac{4\pi^3T_c^3}{7\zeta(3)}
\int_{0}^{\infty}d{\zeta}e^{-\zeta}
\sum_n\frac{1}{(\sqrt{\epsilon_n^2+\zeta W^2})^3}
+1-\beta_c ,
\label{kc}
\end{equation}
where $\beta_c=\beta_a$ (Cf.(\ref{betta})). Appropriate dependencies of $K_C$ 
on $W/T_{c0}$ for the case of $s$-wave pairing are shown in Fig.9.
In case of $d$-wave pairing the picture is similar, but again all the changes
of the coefficient $K_C$ take place on smaller scales of $W/T_{c0}$.

Calculation of the fourth order term in Ginzburg--Landau expansion is 
technically much more complicated. To obtain the expression for the 
coefficient $B$ we have to calculate the trace of the product of four Green's
functions $\hat {\bf G_p}$ which are Nambu matrices of normal and anomalous
Green' functions (\ref{funkgr}):  
\begin{displaymath} 
\hat {\bf G_p}= \left( 
\begin{array}{cc} G_{pp} & F_{pp-Q} \\ F_{p-Qp} & G_{p-Qp-Q} \end{array} 
\right). 
\end{displaymath} 
After performing the trace of $\hat {\bf G_p}\hat 
{\bf G_{-p}}\hat {\bf G_p}\hat {\bf G_{-p}}$ the coefficient $B$ takes the
form:
\begin{eqnarray} 
B=N(0)T_c\sum_{\epsilon_n}\int 
\limits_{0}^{\infty}d{\zeta}e^{-\zeta} \int 
\limits_{0}^{\infty}\frac{d^2{p}}{(2\pi)^2}e^4(\phi) \Biggl\{ 
\left(G_{p,p}G_{-p,-p}+F_{p,p-Q}F_{-p,-p+Q}\right)^2+ \nonumber \\
G_{p,p}G_{-p,-p}F_{-p+Q,p}F_{p-Q,p}+G_{-p+Q,-p+Q}G_{-p,-p}F_{p,p-Q}F_{p-Q,p}+
\nonumber \\
G_{p,p}G_{p-Q,p-Q}F_{-p+Q,-p}F_{-p,-p+Q}+
G_{p-Q,p-Q}G_{-p+Q,-p+Q}F_{p,p-Q}F_{-p,-p+Q} \Biggr \}.
\label{urB}
\end{eqnarray}
It can be seen by direct calculations that the sum of the last four terms in
(\ref{urB}) gives zero, so that:
\begin{equation}
B=N(0)T_c\sum_{\epsilon_n}\int \limits_{0}^{\infty}d{\zeta}e^{-\zeta}
\int \limits_{0}^{\infty}\frac{d^2{p}}{(2\pi)^2}e^4(\phi)
\left(G_{p,p}G_{-p,-p}+F_{p,p-Q}F_{-p,-p+Q}\right)^2.
\label{BBB}
\end{equation}
From here we obtain:
\begin{equation}
B=\frac{N(0)T_c}{2\pi}\sum_n\int \limits_{0}^{\infty}d{\zeta}e^{-\zeta}
\int \limits_{-\infty}^{\infty}d{\xi}\int \limits_{0}^{2\pi}d{\phi}
\frac{e^4(\phi)}{(\epsilon_n^2+\xi^2+\zeta W(\phi)^2)^2},
\label{25}
\end{equation}
and after the integration over $\xi$ and $\phi$ we obtain $K_B$ in the form
similar to (\ref{kc}):
\begin{equation}
K_B=\beta_b\frac{4\pi^3T_c^3}{7\zeta(3)}
\int_{0}^{\infty}d{\zeta}e^{-\zeta}
\sum_n\frac{1}{(\sqrt{\epsilon_n^2+\zeta W^2})^3}
+1-\beta_b,
\label{KB}
\end{equation}
where 
\begin{equation}
\beta_b=\left\{
\begin{array}{ll}
\frac{4\alpha}{\pi} & ,\mbox{$s$-wave pairing} \\
\frac{4\alpha}{\pi}+\frac{4sin(4\alpha)}{3\pi}+\frac{sin(8\alpha)}{6\pi} & ,
\mbox{$d$-wave pairing}
\end{array}.
\right.
\label{bbbb}
\end{equation}
Thus for the case of $s$-wave pairing coefficients $K_B$ and $K_C$ just
coincide.

In conclusion of this section we shall give explicit expressions for
dimensionless Ginzburg--Landau coefficients for the case of $d$-wave pairing
in the model of commensurate fluctuations of short-range order:
\begin{equation}
K_A=\beta_a\frac{1}{2T_c}
\int_{0}^{\infty}d{\zeta}\frac{1}{2\sqrt{\pi\zeta}}e^{-\zeta/4}
\int_{0}^{\omega_c}d{\xi}
\frac{1}{ch^2\frac{\sqrt{\xi^2+\zeta W^2}}{2T_c}}+
1-\beta_a ,
\label{KA}
\end{equation}

\begin{equation}
K_{C,B}=\beta_{c,b}\frac{4\pi^3T_c^3}{7\zeta(3)}
\int_{0}^{\infty}d{\zeta}\frac{1}{2\sqrt{\pi\zeta}}e^{-\zeta/4}
\sum_n\frac{1}{(\sqrt{\epsilon_n^2+\zeta W^2})^3}
+1-\beta_{c,b} ,
\label{18}
\end{equation}
where 
\begin{equation}
\beta_a=\beta_c=\frac{4\alpha-sin(4\alpha)}{\pi}; \qquad 
\beta_b=\frac{4\alpha}{\pi}-\frac{sin(4\alpha)}{6\pi}(5+cos(4\alpha)).
\label{bea}
\end{equation}
It is a simple task to give the appropriate dependencies of these coefficients
on $W/T_{c0}$ for different values of $\alpha$. Qualitatively these 
dependencies are more or less similar to that obtained in incommensurate case,
while all the differences are mainly due to another (larger) scale on the
$W/T_{c0}$ axis(Cf. Fig.6). 

\section{Physical Properties of Superconductors with Pseudogap.}

It is well known that Ginzburg--Landau expansion defines two characteristic
lengths of superconducting state:\ the coherence length and penetration depth
of magnetic field.

The coherence length for given temperature $\xi(T)$ determines characteristic
scale of inhomogeneities of the order parameter $\Delta$,\ i.e. in fact the
``size'' of Cooper's pair:
\begin{equation}
\xi^2(T)=-\frac{C}{A}.
\label{xii}
\end{equation}
In the usual case (in the absence of pseudogap):
\begin{eqnarray}
\xi_{BCS}^2(T)=-\frac{C_{0}}{A_{0}}, \\
\xi_{BCS}(T)\approx 0.74\frac{\xi_{0}}{\sqrt{1-T/T_{c}}},
\label{xiii}
\end{eqnarray}
where $\xi_{0}=0.18v_{F}/T_{c}$.\ In our case:
\begin{equation}
\frac{\xi^2(T)}{\xi_{BCS}^2(T)}=\frac{K_{C}}{K_{A}}.
\label{KCKA}
\end{equation}
Appropriate dependencies of $\xi^2(T)/\xi_{BCS}^2(T)$ on $W/T_{c0}$ for the
case of $d$-wave pairing and incommensurate fluctuations of short range order
are shown in Fig.10.

For the penetration depth in usual case we have:
\begin{equation}
\lambda_{BCS}(T)=\frac{1}{\sqrt{2}}\frac{\lambda_{0}}{\sqrt{1-T/T_{c}}},
\label{lamb}
\end{equation}
where $\lambda_{0}^2=\frac{mc^2}{4\pi ne^2}$ defines penetration depth
for $T=0$.\ In general case we have the following expression for penetration 
depth via Ginzburg--Landau coefficients:
\begin{equation}
\lambda^2(T)=-\frac{c^2}{32\pi e^2}\frac{B}{AC}.
\label{lam}
\end{equation}
Thus for our model:
\begin{equation}
\frac{\lambda(T)}{\lambda_{BCS}(T)}=
\left(\frac{K_{B}}{K_{A}K_{C}}\right)^{1/2}.
\label{lm}
\end{equation}
Graphic dependencies of this parameter on the effective width of the pseudogap
for the case of $d$-wave pairing are shown in Fig.11.

Consider now Ginzburg--Landau parameter:  
\begin{equation} 
\kappa=\frac{\lambda(T)}{\xi(T)}=\frac{c}{4eC}\sqrt{B/2\pi}.
\label{kap}
\end{equation}
In our model of superconductor:
\begin{equation}
\frac{\kappa}{\kappa_{BCS}}=\frac{\sqrt{K_{B}}}{K_{C}},
\label{kp}
\end{equation}
where
\begin{equation}
\kappa_{BCS}=\frac{3c}{\sqrt{7\zeta(3)}e}\frac{T_{c}}{v_{F}^2\sqrt{N(0)}}
\label{kapp}
\end{equation}
--is Ginzburg--Landau parameter for the usual case (without pseudogap).
The dependencies of $\kappa/\kappa_{BCS}$ on $W/T_{c0}$ for the case of
$d$-wave pairing are shown in Fig.12.

Close to $T_{c}$ the upper critical magnetic field $H_{c2}$ is defined by
Ginzburg--Landau coefficients as:
\begin{equation} 
H_{c2}= -\frac{\phi_{0}}{2\pi}\frac{A}{C} ,
\label{12} 
\end{equation} 
where 
$\phi_{0}=c\pi/e$ -- magnetic flux quantum. Then the slope of the upper
critical field at $T_{c}$ is given by:  
\begin{equation} 
\left|\frac{dH_{c2}}{dT}\right|_{T_c}=\frac{24\pi\phi_{0}}{7\zeta(3)v_F^2}T_{c}
\frac{K_A}{K_C}. 
\label{13}
\end{equation}

Graphic dependencies of this slope $\left|\frac{dH_{c2}}{dT}\right|_{T_c}$, 
normalized by the slope at $T_{c0}$, on the effective width of the pseudogap
$W/T_{c0}$ for the case of $d$-wave pairing are shown in Fig.13. It is seen
that the slope drops with the growth of the pseudogap. 

We can also calculate the discontinuity of specific heat at superconducting
transition, which is given by:
\begin{equation} 
\frac{C_s-C_n}{\Omega}=\frac{T_c}{B}\left(\frac{A}{T-T_c}\right)^2,
\label{Cs}
\end{equation}
where $C_s,\>C_n$ -- are specific heats of superconducting and normal states,
$\Omega$ - is system volume. From here the specific heat discontinuity at
$T_{c0}$ ($W=0$) is:
\begin{equation}
\left(\frac{C_s-C_n}{\Omega}\right)_{T_{c0}}=N(0)\frac{8\pi^2T_{c0}}{7\zeta(3)}.
\label{CsCn}
\end{equation}
Then the specific heat discontinuity in our model can be expressed via
dimensionless coefficients $K_A$ and $K_B$ as:
\begin{equation}
\frac{(C_s-C_n)_{T_c}}{(C_s-C_n)_{T_{c0}}}=
\frac{T_c}{T_{c0}}\frac{K_A^2}{K_B}.
\label{cscn}
\end{equation}
The appropriate dependencies on the effective width of the pseudogap for the
case of $d$-wave pairing are shown in Fig.14. It is seen that specific heat
discontinuity is significantly suppressed with the growth of the pseudogap.

Analogous dependencies of all these physical properties for the case of
$s$-wave pairing, and also for the model of commensurate fluctuations, 
qualitatively are more or less similar to that shown in Figs.10-14, the only
major difference is in the scale of $W/T_{c0}$ as in Figs.5-6.

\section{Conclusion.}

In this paper we have analyzed an oversimplified (toy) model of the pseudogap
in two-dimensional electronic system, which is however in some respects
corresponding to some of the observed anomalies of electronic structure of
underdoped HTSC-systems. In particular, it is rather easy to obtain the
``$d$-wave'' symmetry of the pseudogap state, which is determined by the
appropriate positions of ``hot patches'' on the Fermi surface, and strong
scattering by fluctuations of short-range (e.g. antiferromagnetic) order.
Obviously this model can be easily generalized to the case of larger number
of ``hot patches'', so that it can be made more similar to the model of
``hot spots'' \cite{Sch,SchP}. 

The main simplifying assumption as well as the main defect of this model is
our use of rather unrealistic limit of $\xi\rightarrow\infty$ for the
correlation length of fluctuations, which actually helps us to get all the
main results in more or less analytic form. In real systems the value of
$\xi$ is not very large and depends on temperature and doping, becoming an
important independent parameter controlling the physical picture.
In principle our model can be generalized for the case of finite $\xi$ 
along the lines of Refs.\cite{C79,C91}. However, all calculations in this case
become much more complicated. At the same time it is clear enough that the
effect of finite $\xi$ reduces mainly to the smearing of the pseudogap 
\cite{C79,C91} and in this sense is somehow modelled by diminishing the size of
``hot patches'' in our model. It is more or less correct for the effects
determined mainly by the density of states (e.g. superconducting transition 
temperature $T_c$). At the same time this is hardly so in case of the
physical properties, defined mainly by two-particle Green's function, such as
the coefficient $C$ of the gradient term in Ginzburg--Landau expansion. 

Another radical simplification of our model is an assumption of static (and
Gaussian) nature of fluctuations of short range order. This assumption can be
justified in the limit of high enough temperatures $T\gg \omega_{sf}$ 
(where $\omega_{sf}$ - is characteristic frequency of spin fluctuations)
\cite{Pines,Sch,SchP}. Accordingly the assumption of the static nature of these
fluctuations is rather problematic at temperatures close to $T_c$.
However, our analysis of Ginzburg--Landau expansion for different types of
pairing is apparently a rather good description of the main effect of the
``destruction'' of certain parts of the Fermi surface on the main properties
of a superconductor with the pseudogap. It demonstrates an important role of
pseudogap anomalies in formation of superconducting phase for those part of
the phase diagram of HTSC-systems, where pseudogap anomalies are observed 
already in normal state. Analysis of more realistic models is supposed to
be done in future studies.

The authors are grateful to E.Z.Kuchinskii for useful discussions.

This work was performed with partial support from Russian Basic Research
Foundation under the grant No.96-02-16065, as well as under the projects
No.IX.1 of State Program ``Statistical Physics'' and No.96-051 of State
Program on HTSC of the Russian Ministry of Science.

\newpage
\begin{center}
{\bf Figure Captions:}
\end{center}

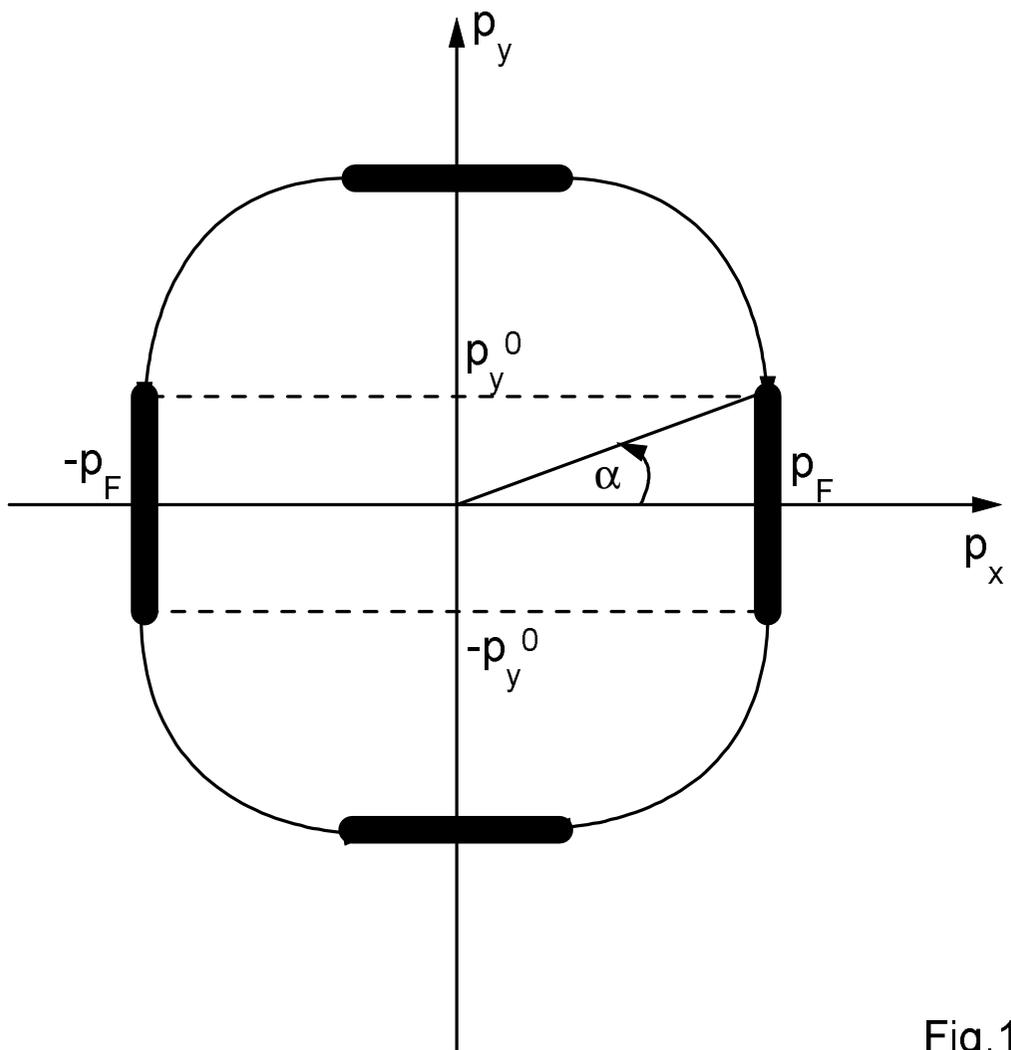
Fig.1. Fermi surface of two-dimensional system. ``Hot patches'' are shown by 
thick lines, with width of the order of $\sim \xi^{-1}$. 

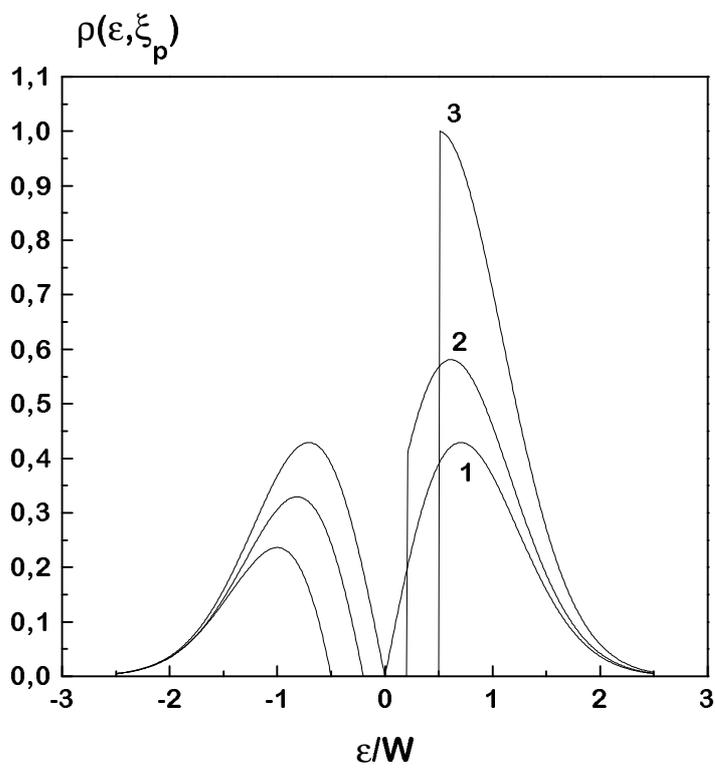
Fig.2. Spectral density of Green's function on the ``hot patch'' of the Fermi
surface.

(1)---$\xi_p=0$;\ (2)---$\xi_p=0.1W$;\ (3)---$\xi_p=0.5W$.

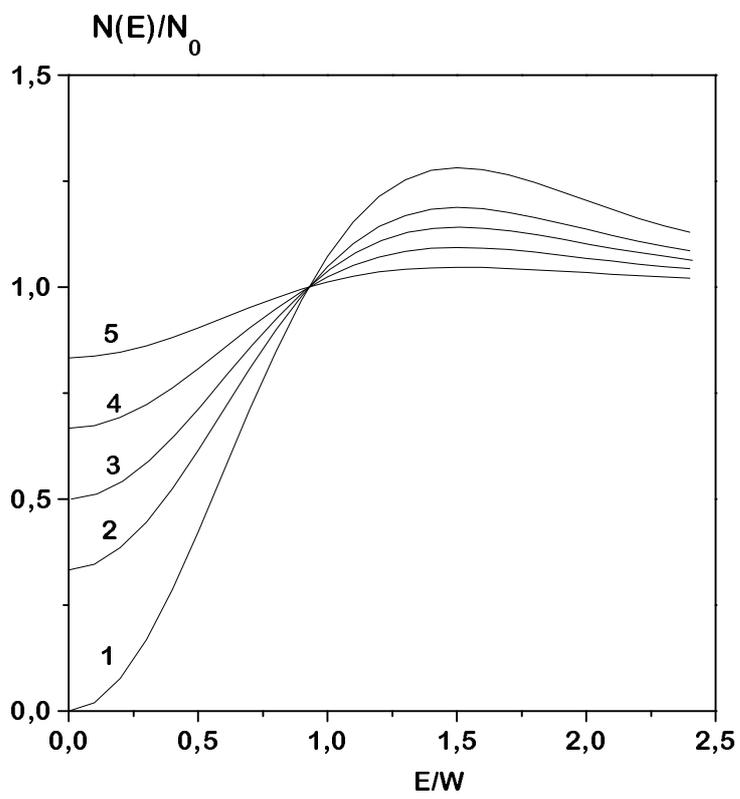
Fig.3. Electronic density of states for ``hot patches'' of different sizes:

(1)---$\alpha=\pi/4$;\ (2)---$\alpha=\pi/6$;\ (3)---$\alpha=\pi/8$;\ 
(4)---$\alpha=\pi/12$;\ (5)---$\alpha=\pi/24$.

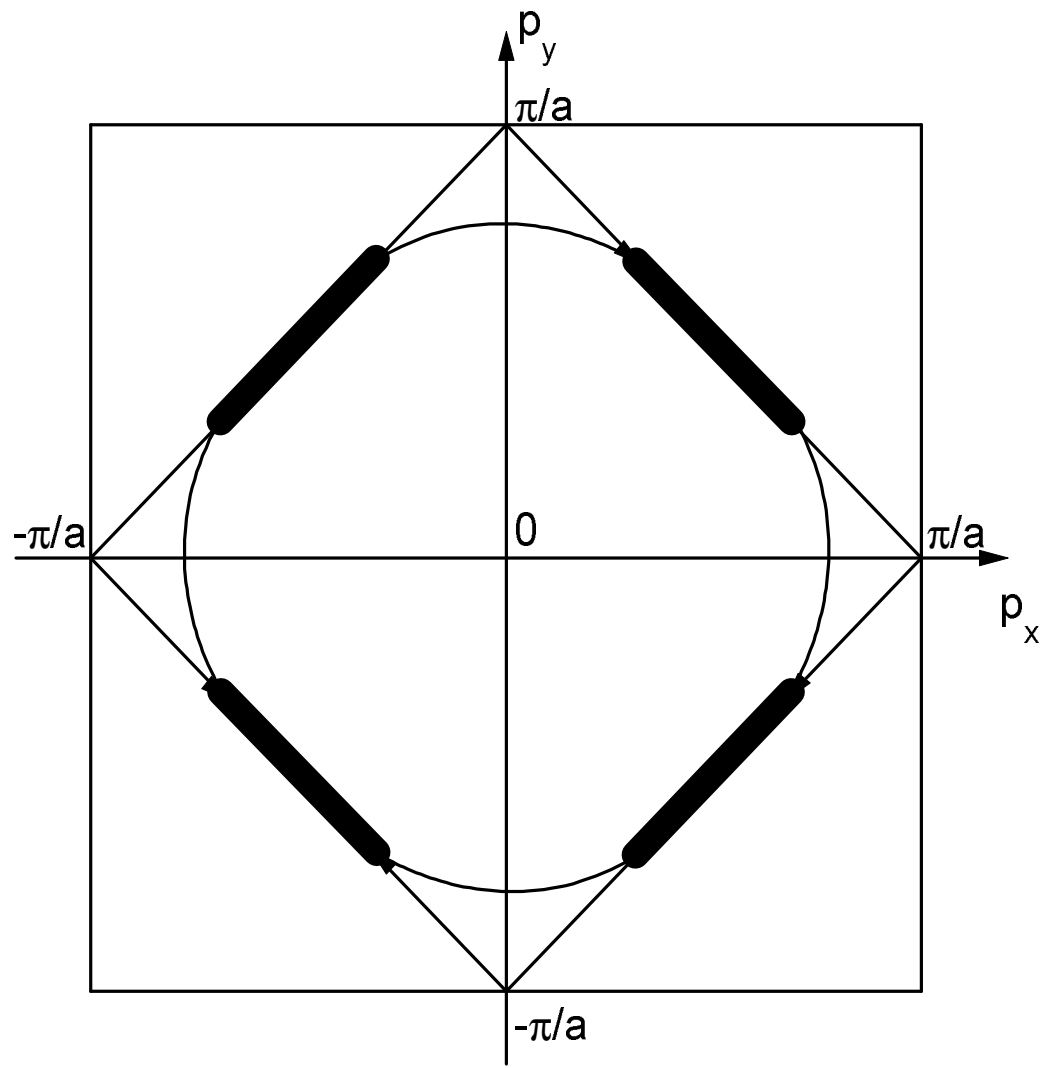
Fig.4. Fermi surface in the Brillouin zone of two-dimensional system in 
``hor patches'' model for the case of commensurate fluctuations of short range
order, corresponding to period doubling. A new Brillouin zone appearing after
the establishment of long range (e.g. antiferromagnetic) order is also shown.

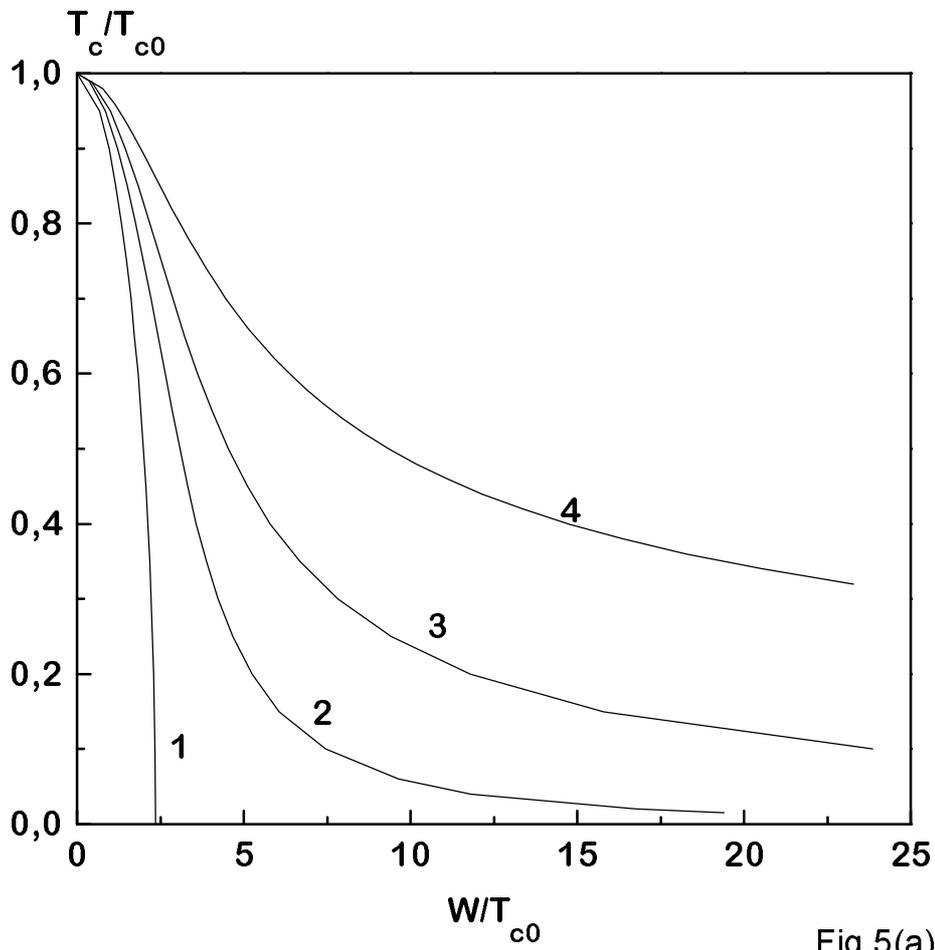
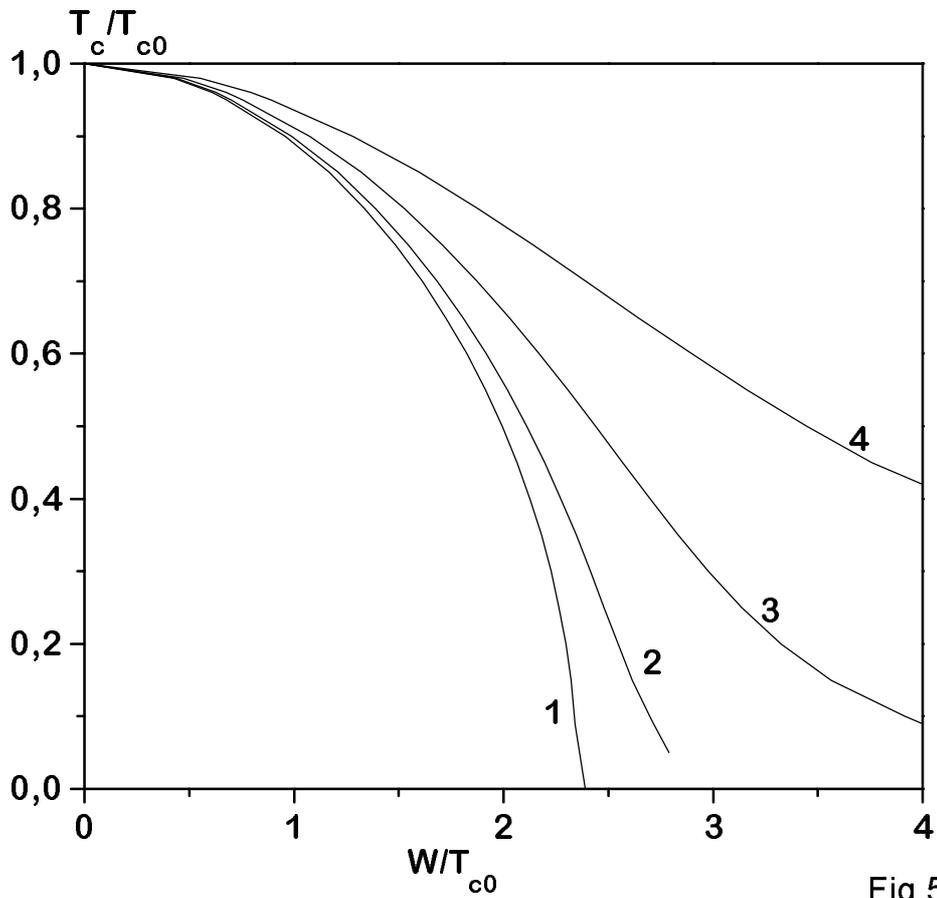
Fig.5.  $T_{c}/T_{c0}$ dependence on the effective width of the pseudogap 
$W/T_{c0}$ for ``hot patches'' of different sizes in the model of
incommensurate fluctuations.

(a)---$s$-wave pairing:
(1)---$\alpha=\pi/4$;\ (2)---$\alpha=\pi/6$;\ (3)---$\alpha=\pi/8$;\ 
(4)---$\alpha=\pi/12$.

(b)---$d$-wave pairing:
(1)---$\alpha=\pi/4$;\ (2)---$\alpha=\pi/6$;\ (3)---$\alpha=\pi/8$;\ 
(4)---$\alpha=\pi/12$.

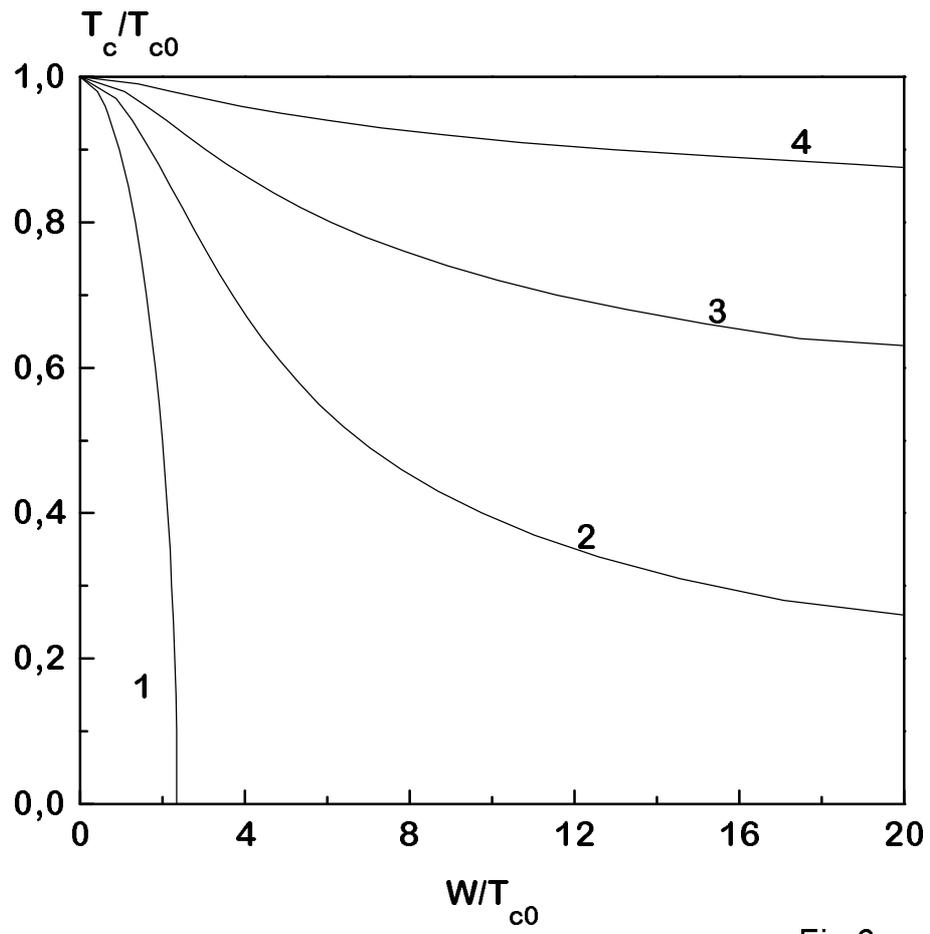
Fig.6. $T_{c}/T_{c0}$ dependence on the effective width of the pseudogap 
$W/T_{c0}$ for ``hot patches'' of different sizes in the model of 
commensurate fluctuations for the case of $d$-wave pairing:

(1)---$\alpha=\pi/4$;\ (2)---$\alpha=\pi/6$;\ (3)---$\alpha=\pi/8$;\ 
(4)---$\alpha=\pi/12$.

Fig.7. Diagrammatic representation of Ginzburg--Landau expansion in the
field of fluctuations of short range order. Electronic lines represent
Nambu matrices composed of normal and anomalous Green's functions
(\ref{funkgr}). Loops are averaged over $\zeta$ with distribution (\ref{intg}) 
or (\ref{integ}). Second loop is calculated for $q=0$ and $T=T_{c}$.

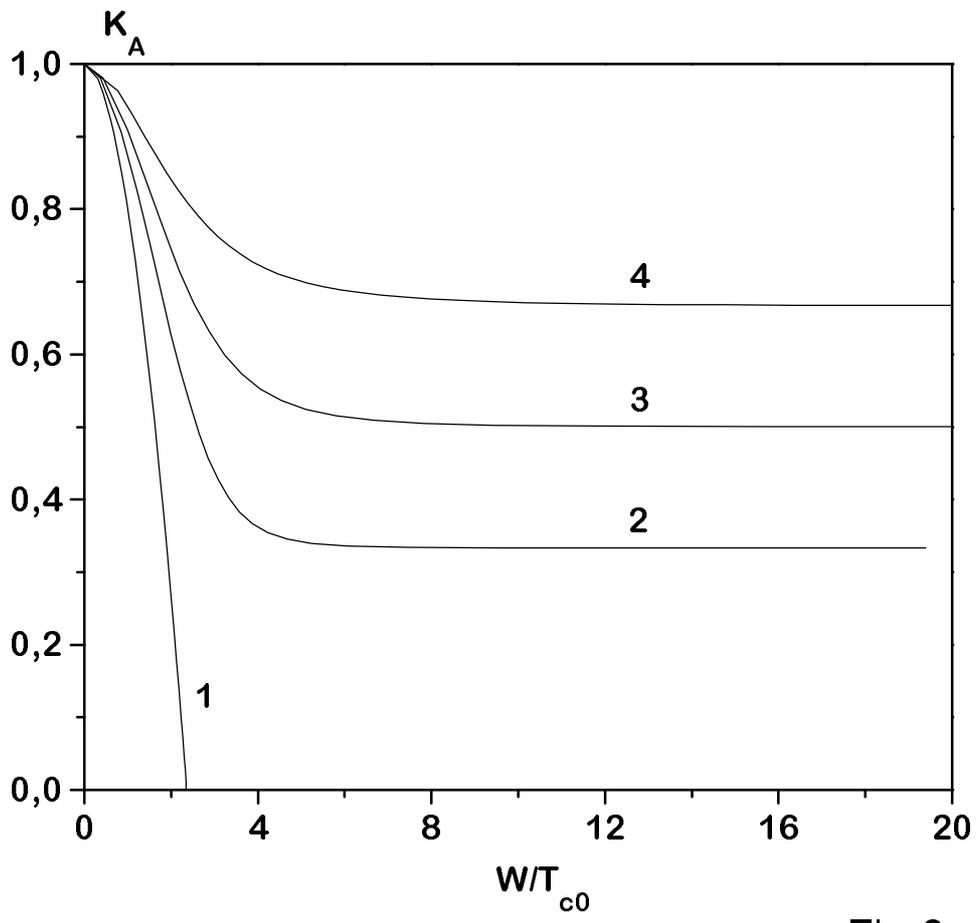
Fig.8. Dependence of the coefficient $K_{A}$ on effective width of the
pseudogap $W/T_{c0}$ for ``hot patches'' of different sizes in the model of
incommensurate fluctuations in the case of $s$-wave pairing:

(1)---$\alpha=\pi/4$;\ (2)---$\alpha=\pi/6$;\ (3)---$\alpha=\pi/8$;\ 
(4)---$\alpha=\pi/12$.

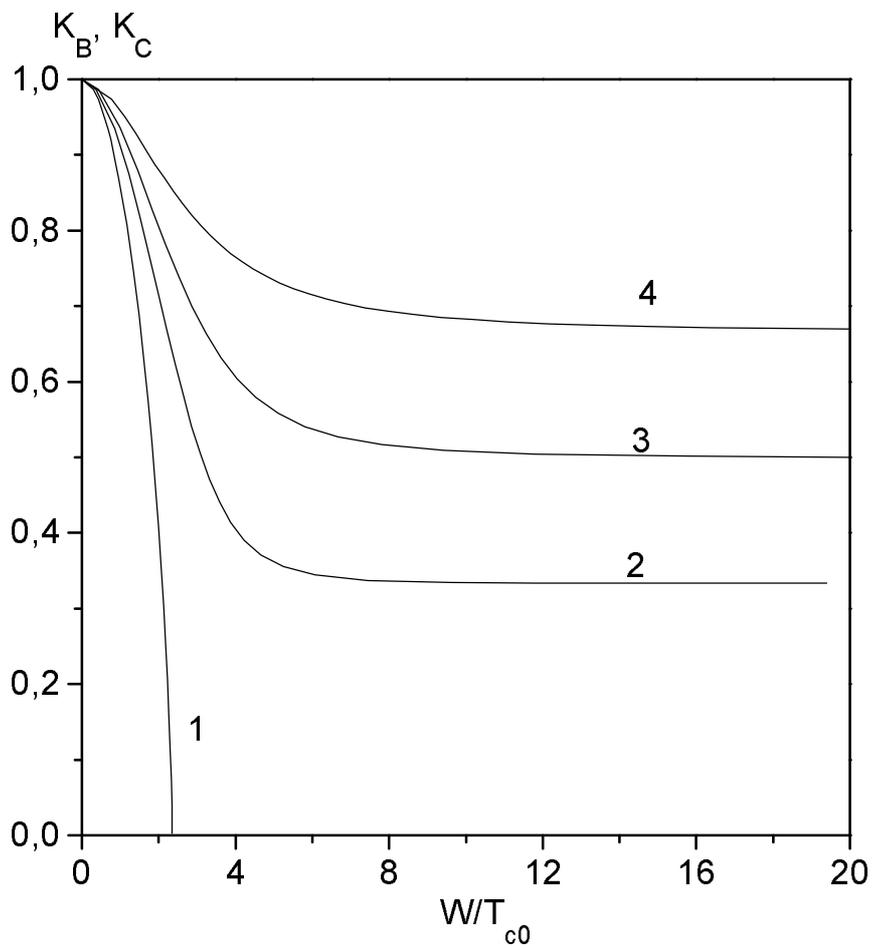
Fig.9. Dependence of coefficients $K_{B}$,\ $K_{C}$ on the effective width of
the pseudogap $W/T_{c0}$ for ``hot patches'' of different sizes in the model
of incommensurate fluctuations in the case of $s$-wave pairing:

(1)---$\alpha=\pi/4$;\ (2)---$\alpha=\pi/6$;\ (3)---$\alpha=\pi/8$;\ 
(4)---$\alpha=\pi/12$.

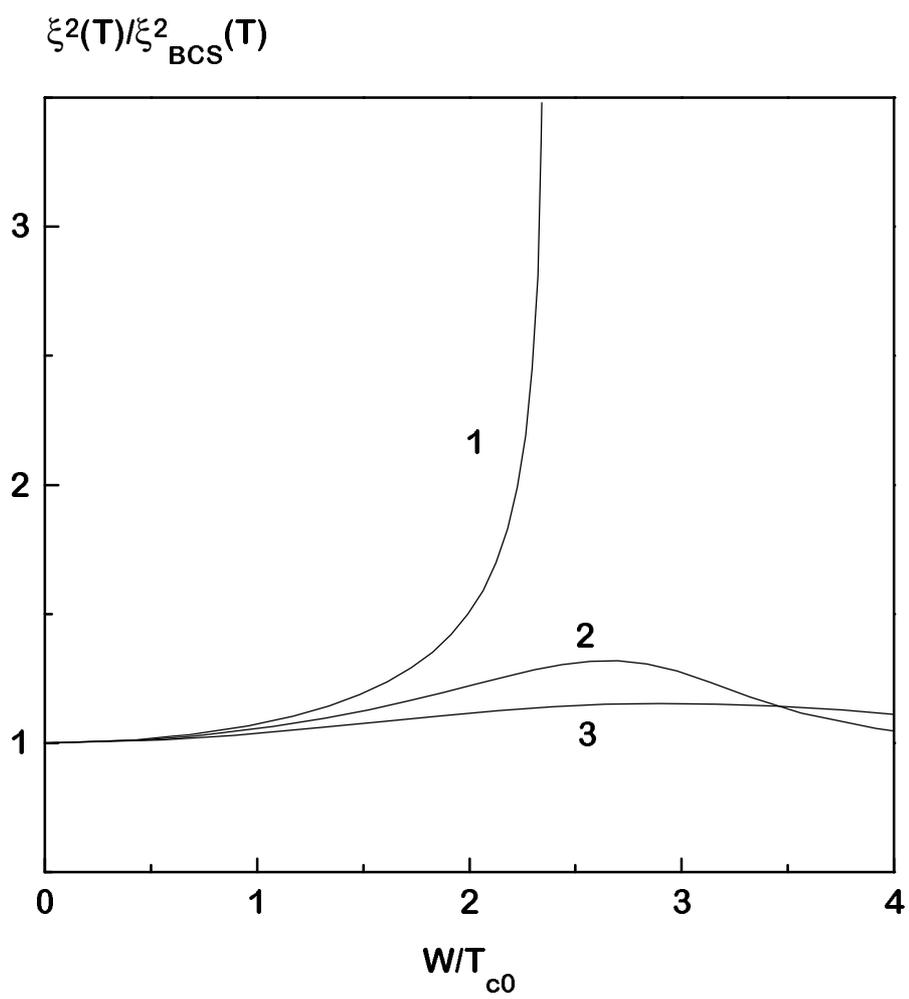
Fig.10. Dependence of coherence length $\xi^2(T)/\xi^2_{BCS}(T)$ on the
effective width of the pseudogap $W/T_{c0}$ in case of $d$-wave pairing:

(1)---$\alpha=\pi/4$;\ (2)---$\alpha=\pi/8$;\ (3)---$\alpha=\pi/12$.

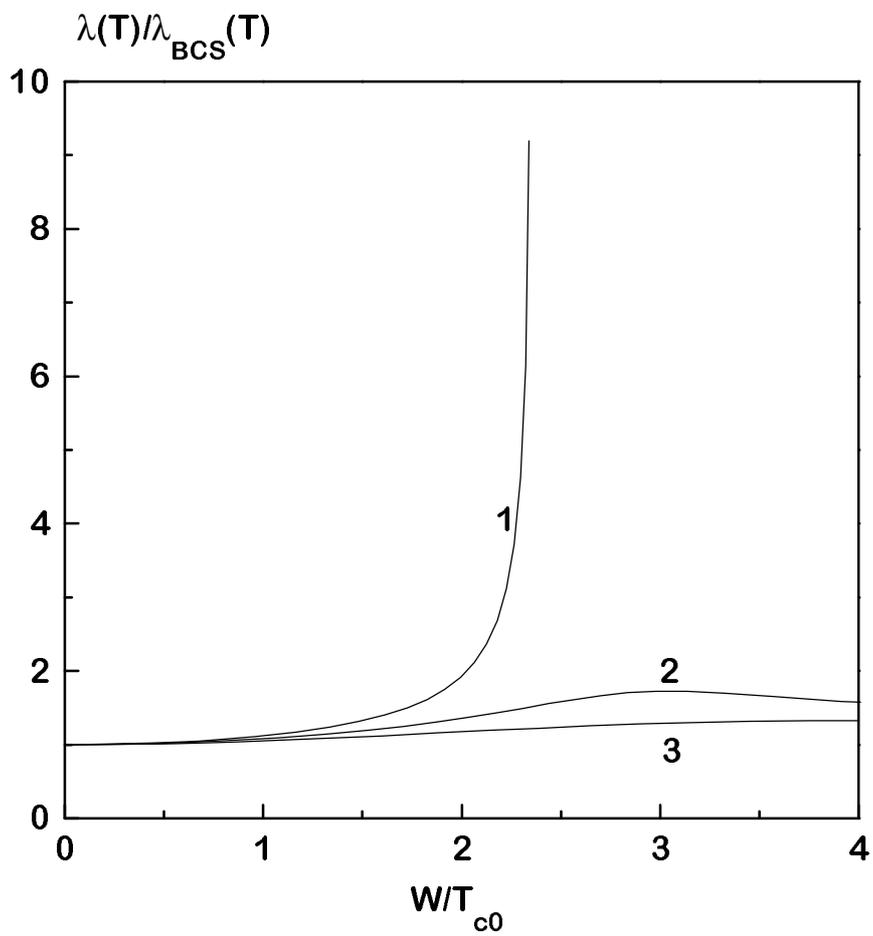
Fig.11. Dependence of penetration depth $\lambda(T)/\lambda_{BCS}(T)$ on the 
effective width of the pseudogap $W/T_{c0}$ in case of $d$-wave pairing:

(1)---$\alpha=\pi/4$;\ (2)---$\alpha=\pi/8$;\ (3)---$\alpha=\pi/12$.

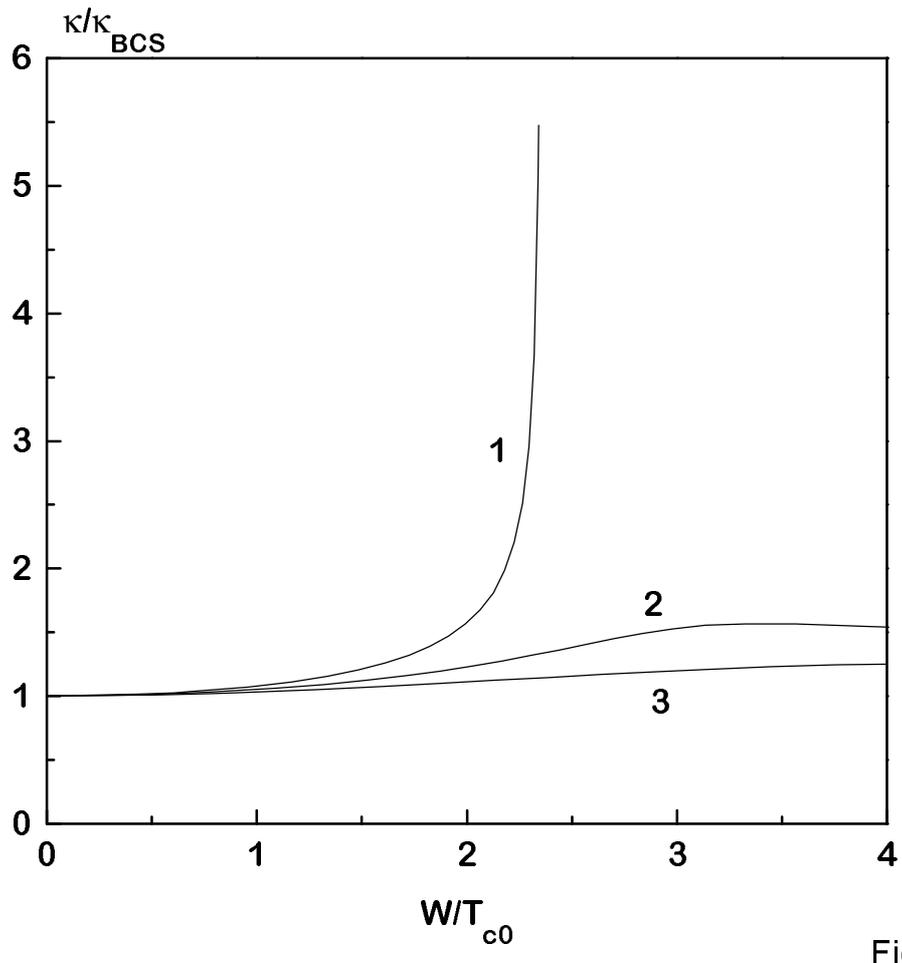
Fig.12. Dependence of Ginzburg--Landau parameter $\kappa/\kappa_{BCS}$ on the
effective width of the pseudogap $W/T_{c0}$ in case of $d$-wave pairing:

(1)---$\alpha=\pi/4$;\ (2)---$\alpha=\pi/8$;\ (3)---$\alpha=\pi/12$.

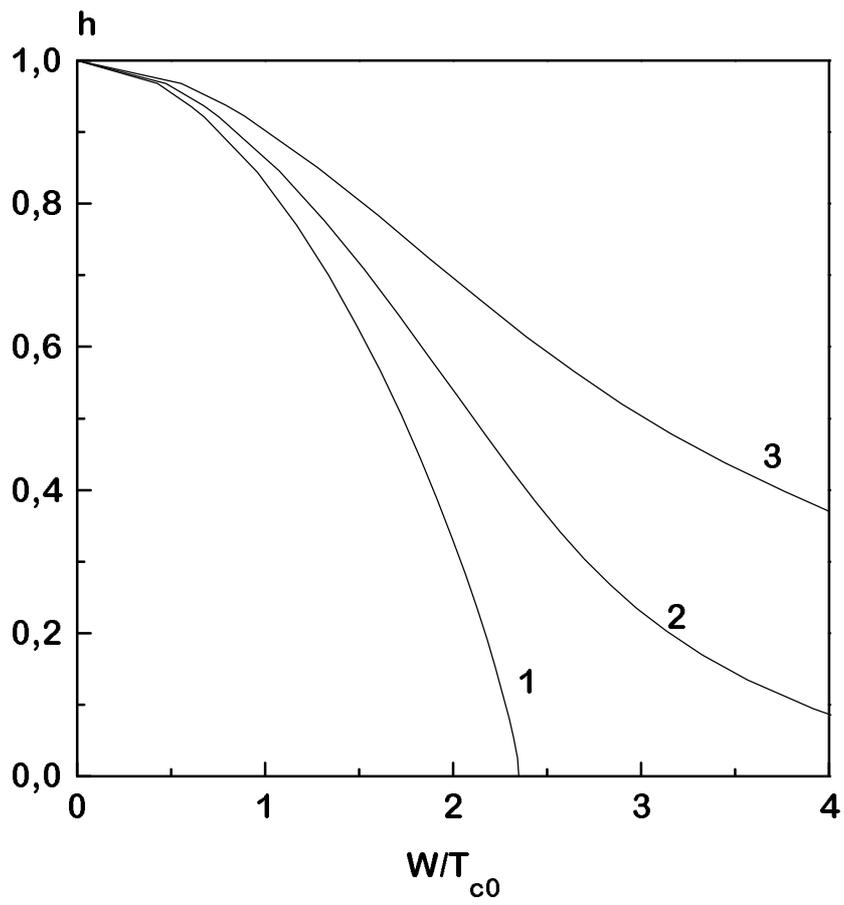
Fig.13. Dependence of normalized slope of the upper critical field
on the effective width of the pseudogap $W/T_{c0}$ in case of $d$-wave
pairing:

(1)---$\alpha=\pi/4$;\ (2)---$\alpha=\pi/8$;\ (3)---$\alpha=\pi/12$.

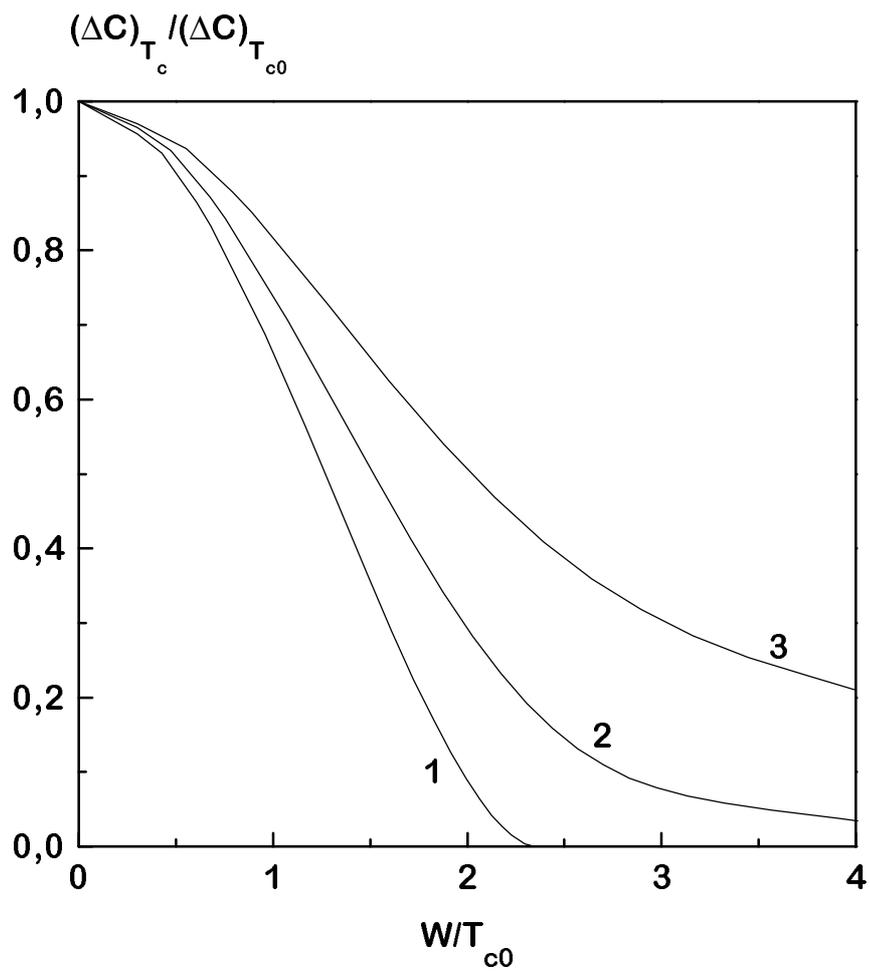
Fig.14. Dependence of normalized discontinuity of specific heat on the
effective width of the pseudogap $W/T_{c0}$ in case of $d$-wave pairing:

(1)---$\alpha=\pi/4$;\ (2)---$\alpha=\pi/8$;\ (3)---$\alpha=\pi/12$.

\end{document}